\documentclass[twocolumn]{revtex4-1}

\usepackage{graphicx}
\usepackage{dcolumn}
\usepackage{amsmath}

\begin{document}

\title{Renormalization group theory of effective field theory models
in low dimensions
}

\author{Takashi Yanagisawa}

\affiliation{
National Institute of Advanced Industrial Science and Technology (AIST),
Tsukuba Central 2, 1-1-1 Umezono, Tsukuba 305-8568, Japan
}


\begin{abstract}
This is a lecture note on the renormalization group theory for field
theory models based on
the dimensional regularization method.
We discuss the renormalization group approach to fundamental field
theoretic models in low dimensions.
We consider the models that are universal and frequently appear in
physics, both in high-energy physics and condensed-matter physics.
They are the non-linear sigma model, the $\phi^4$ model and the 
sine-Gordon model.
We use the dimensional regularization method to regularize the
divergence and derive the renormalization group equations
called the beta functions.  The dimensional method is described
in detail.
\end{abstract}

\maketitle

\section{Introduction}

The renormalization group is a fundamental and powerful tool
to investigate the property of quantum 
systems\cite{zin02,gro76,itz80,col85,pes95,ram89,wei95,ryd85,nas78,
nis69,bog80,and84,par88,ami05,pol87}.
The physics of a many-body system is sometimes captured by
the analysis of an effective field theory 
model\cite{abr65,tsv95,fra91,pop87}.
Typically, effective field-theory models are the $\phi^4$ theory,
the non-linear sigma model and the sine-Gordon model.
Each of these models represents universality as a representative
of a universal class.

The $\phi^4$ theory is the model of a phase transition, which is
often referred to as the Ginzburg-Landau model.
The renormalization of the $\phi^4$ theory gives a prototype of
renormalization group procedures in field 
theory\cite{wil72,wil72b,wil74,bre72,bre82}.

The non-linear sigma model appears in various fields of 
physics\cite{pol87,pol75,bre76,zak89}, and is the effective
model of QCD\cite{ell03} and that of magnets
(ferromagnet and antiferromgnetic materials)\cite{nel77,cha88,cha89,yan92}.
This model exhibits an important property called the
asymptotic freedom.  The non-linear sigma model is
generalized to a model with fields that take values in a
compact Lie group $G$\cite{per89,bre80,hik80a,hik80b,hik83,wes71,wit83,wit84,nov82,gol78}.
This is called the chiral model.

The sine-Gordon model also has universality\cite{col75,bre79,das79,zam79,raj82,man04,yan16b}.
The two-dimensional (2D) sine-Gordon model describes the
Kosterlitz-Thouless transition of the 2D classical XY 
model\cite{kos73,kos74}.
The 2D sine-Gordon model is mapped to the Coulomb gas model
where particles interact with each other through a
logarithmic interaction.  The Kondo problem\cite{kon64,kon12}
also belongs to the same universality class where the
scaling equations are just given by those for the 2D 
sine-Gordon model, that is, the equations for the
Kosterlitz-Thouless transition\cite{kon12,and70,and69,yuv70,and70b}.
The one-dimensional Hubbard model is also mapped onto 
the 2D sine-Gordon model on the basis of a bosonization
method\cite{sol79,hal66}.
The Hubbard model is an important model of strongly
correlated electrons\cite{hub63,yam98,yan01,yan03,yam00,yan16}.
The Nambu-Goldstone (NG) modes in a multi-gap superconductor
becomes massive due to the cosine potential, and thus the
dynamical property of the NG mode can be understood by
using the sine-Gordon model\cite{leg66,tan10,tan10b,yan12,yan13,yan14}.
The sine-Gordon model will play an important role in layered
high-temperature superconductors because the Josephson
plasma oscillation is analysed using this model\cite{kle13,tam92,mat95,koy96}.

In this paper, we discuss the renormalization group
theory for the $\phi^4$ theory, the non-linear sigma model
and the sine-Gordon model.  We use the dimensional
regularization procedure to regularize the divergence\cite{tho72}.

\section{$\phi^4$ Model}

\subsection{Lagrangian}

The $\phi^4$ model is given by the Lagrangian
\begin{equation}
\mathcal{L}= \frac{1}{2}\left(\partial_{\mu}\phi\right)^2
-\frac{1}{2}m^2\phi^2-\frac{1}{4!}g\phi^4,
\end{equation}
where $\phi$ is a real scalar field and $g$ is the coupling
constant.
In the unit of the momentum $\mu$, the dimension of $\mathcal{L}$
is given by $d$ where $d$ is the dimension of the space-time:
$[\mathcal{L}]=\mu^d$.  The dimension of the field $\phi$ is
$(d-2)/2$: $[\phi]=\mu^{(d-2)/2}$.  Because $g\phi^4$ has the
dimension $d$, the dimension of $g$ is given by $4-d$:
$[g]=\mu^{4-d}$.  Let us adopt that $\phi$ has $N$ components:
$\phi=(\phi_1,\phi_2,\cdots,\phi_N)$.  The interaction term
$\phi^4$ is defined as
\begin{equation}
\phi^4= \left( \sum_{i=1}^N\phi_i^2\right)^2.
\end{equation}
The Green's function is defined as
\begin{equation}
G_i(x-y)= -i\langle 0|T\phi_i(x)\phi_i(y)|0\rangle,
\end{equation}
where $T$ is the time-ordering operator and $|0\rangle$ is the
ground state.  The Fourier transform of the Green's
function is
\begin{equation}
G_i(p)= \int d^dx e^{ip\cdot x}G_i(x).
\end{equation}
In the non-interacting case with $g=0$, the Green's function
is given by
\begin{equation}
G_i^{(0)}(p)= \frac{1}{p^2-m^2},
\end{equation}
where $p^2= (p_0)^2-{\bf p}^2$ for $p=(p_0,{\bf p})$.

Let us consider the correction to the Green's function by
means of the perturbation theory in terms of the interaction
term $g\phi^4$.  A diagram that appears in perturbative
expansion contains in general $L$ loops, $I$ internal lines
and $V$ vertices.  They are related by
\begin{equation}
L= I-V+1.
\end{equation}
There are $L$ degrees of freedom for momentum integration.
The degree of divergence $D$ is given by
\begin{equation}
D= d\cdot L-2L.
\end{equation}
We have a logarithmic divergence when $D=0$.  Let $E$ be the
number of external lines.  We obtain
\begin{equation}
4V= E+2I.
\end{equation}
Then the degree of divergence is written as
\begin{equation}
D= d\cdot L-2I= d+(d-4)V+\left(1-\frac{d}{2}\right)E.
\end{equation}
In four dimensions, the degree of divergence $D$ is independent of
the numbers of internal lines and vertices:
\begin{equation}
D= 4-E.
\end{equation}
When the diagram has four external lines, $E=4$, we obtain $D=0$
which indicates that we have a logarithmic (zero-order) divergence.
This divergence can be renormalized.

Let us consider the Lagrangian with bare quantities:
\begin{equation}
\mathcal{L}= \frac{1}{2}(\partial_{\mu}\phi_0)^2
-\frac{1}{2}m_0^2\phi_0^2-\frac{1}{4!}g_0\phi_0^4,
\end{equation}
where $\phi_0$ denotes the bare field, $g_0$ denotes
the bare coupling constant and $m_0$ is the bare mass.
We introduce the renormalized field $\phi$, the renormalized
coupling constant $g$ and the renormalized mass $m$.
They are defined by
\begin{align}
\phi_0 &= \sqrt{Z_{\phi}}\phi,\\
g_0 &= Z_{g}g,\\
m_0^2 &= m^2 Z_{2}/Z_{\phi},
\end{align}
where $Z_{\phi}$, $Z_g$ and $Z_2$ are renormalization constants.
When we write $Z_g$ as
\begin{equation}
Z_g = Z_4/Z_{\phi}^2,
\end{equation}
we have $g_0Z_{\phi}^2= gZ_4$.  Then the Lagrangian is
written by means of renormalized field and constants:
\begin{equation}
\mathcal{L}= \frac{1}{2}Z_{\phi}(\partial_{\mu}\phi)^2
-\frac{1}{2}m^2Z_2\phi^2 -\frac{1}{4!}gZ_4\phi^4.
\end{equation}

\subsection{Regularization of divergences}

\subsubsection{Two-point function}

We use the perturbation theory in terms of the interaction
$g\phi^4$.  For a multi-component scalar field theory,
it is convenient to express the interaction $\phi^4$
as in Fig.1 where the dashed line indicates the coupling $g$.
We first examine the massless case with $m\rightarrow 0$.
Let us consider the renormalization of the two-point function
$\Gamma^{(2)}(p)= iG(p)^{-1}$.
The contributions to $\Gamma^{(2)}$ are shown in Fig.2.
The first term indicates $p^2Z_{\phi}$ and the contribution
in the second term is represented by the integral
\begin{equation}
I := \int \frac{d^dq}{(2\pi)^d}\frac{1}{q^2-m^2}.
\end{equation}
Using the Euclidean coordinate $q_4= -iq_0$, this integral
is evaluated as
\begin{equation}
I= -i\frac{\Omega_d}{(2\pi)^d}m^{d-2}\frac{1}{2}
\Gamma\left(\frac{d}{2}\right)\Gamma\left(1-\frac{d}{2}\right),
\end{equation}
where $\Omega_d$ is the solid angle in $d$ dimensions.
For $d>2$, the integral $I$ vanishes in the limit
$m\rightarrow 0$.  Thus the mass remains zero in the massless
case.  We do not consider mass renormalization in the massless
case.  

Let us examine the third term in Fig.2.
There are $4^2\cdot 2N+4^2\cdot 2^2=32N+64$ ways to connect lines
for an $N$-component scalar field to form the third diagram
in Fig.2.  This is seen by noticing that this diagram is
represented as a sum of two terms in Fig.3.
The number of ways to connect lines is $32N$ for (a) and 64
for (b).  Then we have the factor from these contributions as
\begin{equation}
\left( \frac{1}{4!}g\right)^2 (32N+64)= \frac{N+2}{18}g^2.
\end{equation}

The momentum integral of this term is given as
\begin{equation}
J(k) := \int \frac{d^dp}{(2\pi)^d}\frac{d^dq}{(2\pi)^d}
\frac{1}{p^2q^2(p+q+k)^2}.
\end{equation}
The integral $J$ exhibits a divergence in four dimensions
$d=4$.  We separate the divergence as $1/\epsilon$ by
adopting $d=4-\epsilon$.  The divergent part is regularized as
\begin{equation}
J(k)- -\left(\frac{1}{8\pi^2}\right)^2 \frac{1}{8\epsilon}k^2
+{\rm regular}~~{\rm terms}.
\end{equation}
To obtain this, we first perform the integral with respect 
to $q$ by using
\begin{equation}
\frac{1}{q^2(p+q+k)^2}= \int_0^1 dx
\frac{1}{[q^2x+(p+q+k)^2(1-x)]^2}.
\end{equation}
For $q'=q+(1-x)(p+k)$, we have
\begin{eqnarray}
&&\int\frac{d^dq}{(2\pi)^d}\frac{1}{q^2(p+q+k)^2}\nonumber\\
&&= \int\frac{d^dq'}{(2\pi)^d}\int_0^1 dx
\frac{1}{[q'^2+x(1-x)(p+k)^2]^2} \nonumber\\
&&= \frac{\Omega_d}{(2\pi)^d}\frac{1}{2}
\Gamma\left(\frac{d}{2}\right)\Gamma\left(2-\frac{d}{2}\right)
\Gamma\left(\frac{d}{2}-1\right)^2\frac{1}{\Gamma(d-2)}
\nonumber\\
&&\times \left( (p+k)^2\right)^{\frac{d}{2}-2}.
\end{eqnarray}
Here the following parameter formula was used:
\begin{equation}
\frac{1}{A^nB^m}= \frac{\Gamma(n+m)}{\Gamma(n)\Gamma(m)}
\int_0^1 dx \frac{x^{n-1}(1-x)^{m-1}}{[xA+(1-x)B]^{n+m}}.
\end{equation}
Then, we obtain
\begin{eqnarray}
&& \int\frac{d^dp}{(2\pi)^d}\frac{1}{p^2((p+k)^2)^{2-d/2}}
\nonumber\\
&&= \frac{\Gamma(3-d/2)}{\Gamma(2-d/2)}\int_0^1 dx(1-x)^{1-d/2}
\nonumber\\
&&\times \int\frac{d^dp'}{(2\pi)^d}\frac{1}{[p'^2+x(1-x)k^2]^{3-d/2}}
\nonumber\\
&&= \frac{\Omega_d}{(2\pi)^d}\frac{\Gamma(3-d/2)}{\Gamma(2-d/2)}
B\left(d-2,\frac{d}{2}-1\right) \nonumber\\
&&\times \frac{1}{2}B\left(\frac{d}{2},3-d\right)(k^2)^{d-3}.
\end{eqnarray}
Here $B(p,q)= \Gamma(p)\Gamma(q)/\Gamma(p+q)$.
We use the formula
\begin{equation}
\Gamma(\epsilon)= \frac{1}{\epsilon}+{\rm finite}~~{\rm terms},
\end{equation}
for $\epsilon\rightarrow 0$.  This results in
\begin{eqnarray}
\int\frac{d^dp}{(2\pi)^d}\int\frac{d^dq}{(2\pi)^d}
\frac{1}{p^2q^2(p+q+k)^2} &=& -\left(\frac{1}{8\pi^2}\right)^2
\frac{1}{8\epsilon}k^2 \nonumber\\
&+&  {\rm regular}~~{\rm terms}.\nonumber\\
\end{eqnarray}
Therefore, the two-point function is evaluated as
\begin{equation}
\Gamma^{(2)}(p)= Z_{\phi}p^2+\frac{1}{8\epsilon}
\frac{N+2}{18}\left(\frac{g}{8\pi^2}\right)^2g^2,
\end{equation}
up to the order of $O(g^2)$. 
In order to cancel the divergence, we choose $Z_{\phi}$ as
\begin{equation}
Z_{\phi}= 1-\frac{1}{8\epsilon}\frac{N+2}{18}
\left(\frac{1}{8\pi^2}\right)^2g^2.
\end{equation}

\begin{figure}[htbp]
\begin{center}
  \includegraphics[height=2.5cm]{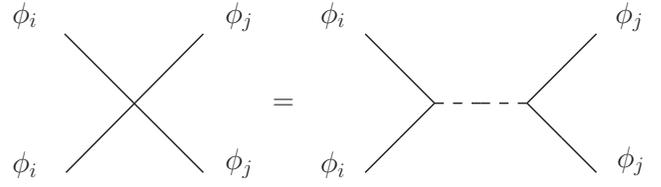}
\caption{
$\phi^4$ interaction with the coupling constant $g$.
}
\label{fig1}       
\end{center}
\end{figure}

\begin{figure}[htbp]
\begin{center}
  \includegraphics[width=9.2cm]{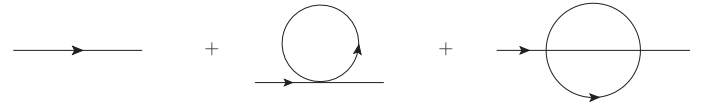}
\caption{
The contributions to the two-point function $\Gamma^{(2)}(p)$
up to the order of $g^2$.
}
\label{fig2}       
\end{center}
\end{figure}

\begin{figure}[htbp]
\begin{center}
  \includegraphics[height=2.5cm]{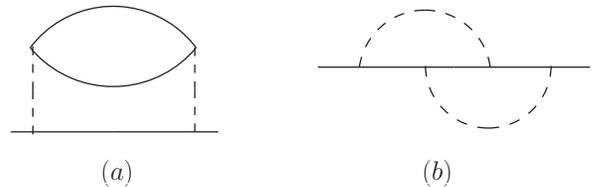}
\caption{
The third term in Fig.2 is a sum of two configurations
in (a) and (b).
}
\label{fig3}       
\end{center}
\end{figure}

\subsubsection{Four-point function}

Let us turn to the renormalization of the interaction term $g^4$.
The perturbative expansions of the four-point function is shown
in Fig.4.  The diagram (b) in Fig.4, denoted as 
$\Delta\Gamma^{(4)}_b$, is given by for $N=1$:
\begin{equation}
\Delta\Gamma^{(4)}_b(p)= g^2\frac{1}{2}\int\frac{d^dq}{(2\pi)^d}
\frac{1}{(q^2-m^2)((p+q)^2-m^2)}.
\end{equation}
As in the calculation of the two-point function, this is
regularized as
\begin{equation}
\Delta\Gamma^{(4)}_b(p)= i\frac{1}{8\pi^2}\frac{1}{2\epsilon}g^2,
\end{equation}
for $d=4-\epsilon$.  Let us evaluate the multiplicity of this
contribution for $N>1$.  For $N=1$, we have a factor
$4^23^22/4!4!=1/2$ as shown in eq.(30).
The diagrams (c) and (d) in Fig.4 give the same contribution
as in eq.(31), giving the factor 3/2 as a sum of (b), (c) and (d).
For $N>1$, there is a summation with respect to the components
of $\phi$.  We have the multiplicity factor for the diagram
in Fig.4(b) as
\begin{equation}
\left( \frac{1}{4!}\right)^2 2^22^22N= \frac{N}{18}.
\end{equation}
Since we obtain the same factor for diagrams in Fig.4(c)
and 4(d), we have $N/6$ in total.  We subtract 1/6 for
$N=1$ from 3/2 to have 8/6.  As a result the multiplicity
factor is given by $(N+8)/6$.  Then, the four-point function
is regularized as
\begin{equation}
\Delta\Gamma^{(4)}(p)= i\frac{1}{8\pi^2}\frac{N+8}{6}
\frac{1}{\epsilon}g^2.
\end{equation}
Because $g$ has the dimension $4-d$ such as $[g]=\mu^{4-d}$,
we write $g$ as $g\mu^{4-d}$ so that $g$ is the dimensionless
coupling constant.  Now we have
\begin{equation}
\Gamma^{(4)}(p)= -igZ_4\mu^{\epsilon}
+i\frac{1}{8\pi^2}\frac{N+8}{6}\frac{1}{\epsilon}g^2,
\end{equation}
for $d=4-\epsilon$ where we neglected $\mu^{\epsilon}$ in the second
term.  The renormalization constant is determined as.
\begin{equation}
Z_4= 1+\frac{N+8}{6\epsilon}\frac{1}{8\pi^2}g.
\end{equation}
As a result, the four-point function $\Gamma^{(4)}$ becomes
finite.

\begin{figure}[htbp]
\begin{center}
  \includegraphics[height=1.5cm]{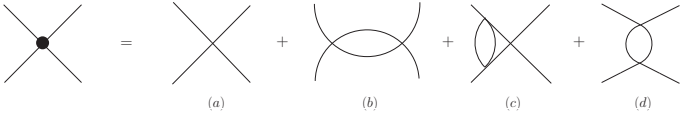}
\caption{
Diagrams for four-point function.
}
\label{fig4}       
\end{center}
\end{figure}

\subsection{Beta function $\beta(g)$}

The bare coupling constant is written as
$g_0= Z_g g\mu^{4-d}=(Z_4/Z_{\phi}^2)g\mu^{4-d}$.  Since $g_0$ is 
independent of the energy scale $\mu$, we have
$\mu\partial g_0/\partial\mu=0$.  This results in
\begin{equation}
\mu\frac{\partial g}{\partial\mu}= (d-4)g
-g\mu\frac{\partial g}{\partial\mu}\frac{\partial\ln Z_g}{\partial g},
\end{equation}
where $Z_g=Z_4/Z_{\phi}^2$.  We define the beta function for $g$ as
\begin{equation}
\beta(g)= \mu\frac{\partial g}{\partial\mu},
\end{equation}
where the derivative is evaluated under the condition that the bare
$g_0$ is fixed.  Because
\begin{equation}
Z_g = 1+\frac{N+1}{6\epsilon}\frac{1}{8\pi^2}g+O(g^2),
\end{equation}
the beta function is given as
\begin{equation}
\beta(g)= \frac{-\epsilon g}{1+g\frac{\partial\ln Z_g}{\partial g}}
= -\epsilon g+\frac{N+8}{6}\frac{1}{8\pi^2}g^2+O(g^3).
\end{equation}
$\beta(g)$ up to the order of $g^2$ is shown as a function of $g$
for $d<4$ in Fig.5.  For $d<4$, there is a non-trivial fixed point at
\begin{equation}
g_c = \epsilon\frac{48\pi^2}{N+8}.
\end{equation}
For $d=4$, we have only a trivial fixed point at $g=0$.

For $d=4$ and $N=1$, the beta function is given by
\begin{equation}
\beta(g)= \frac{3}{16\pi^2}g^2+\cdots.
\end{equation}
In this case, the $\beta(g)$ has been calculated up to the 5th order
of $g$\cite{vla79}:
\begin{eqnarray}
\beta(g) &=& \frac{3}{16\pi^2}g^2-\frac{17}{3}\frac{1}{(16\pi^2)^2}g^3
\nonumber\\
&+& \left(\frac{145}{8}+12\zeta(3)\right)\frac{1}{(16\pi^2)^2}g^4
+ A_5\frac{1}{(16\pi^2)^4}g^5,
\nonumber\\
\end{eqnarray}
where
\begin{equation}
A_5 = -\left( \frac{3499}{48}+78\zeta(3)-18\zeta(4)+120\zeta(5)\right),
\end{equation}
and $\zeta(n)$ is the Riemann zeta function.
The renormalization constant $Z_g$ and the beta function $\beta(g)$
are obtained as a power series of $g$.  We express $Z_g$ as
\begin{equation}
Z_g= 1+\frac{N+8}{6\epsilon}g
+\left(\frac{b_1}{\epsilon^2}+\frac{b_2}{\epsilon}\right)g^2
+\left( \frac{c_1}{\epsilon^3}+\frac{c_2}{\epsilon^2}+\frac{c_3}{\epsilon}
\right)g^3+\cdots,
\end{equation}
and then the $\beta(g)$ is written as
\begin{eqnarray}
\beta(g)&=& -\epsilon g+\epsilon g^2\Big[ \frac{N+8}{6\epsilon}
+2\left(\frac{b_1}{\epsilon^2}+\frac{b_2}{\epsilon}\right)g
\nonumber\\
&& +\frac{(N+8)^2}{36\epsilon^2}g+\cdots\Big]\nonumber\\
&=& -\epsilon g+\frac{N+8}{6}g^2-\frac{9N+42}{36}g^3+\cdots .
\end{eqnarray}
Here the factor $1/8\pi^2$ is included in $g$.
The terms of order $1/\epsilon^2$ are cancelled because
\begin{equation}
b_1= -\frac{(N+8)^2}{72}.
\end{equation}
In general, the $n$-th order term in $\beta(g)$ is given by $n!g^n$.
The function $\beta(g)$ is expected to have the form
\begin{equation}
\beta(g)= -\epsilon g+\frac{N+8}{6}g^2+\cdots
+n!a^nn^bcg^n +\cdots ,
\end{equation}
where a, b and c are constants.

\begin{figure}[htbp]
\begin{center}
  \includegraphics[height=4.2cm]{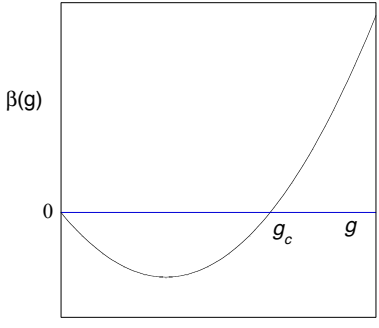}
\caption{
The beta function of $g$ for $d<4$.  There is a finite
fixed point $g_c$.
}
\label{fig5}       
\end{center}
\end{figure}

\subsection{n-point function and anomalous dimension}
 Let us consider the $n$-point function $\Gamma^{(n)}$.
The bare and renormalized $n$-point functions are denoted as
$\Gamma^{(n)}_B(p_i,g_0,m_0^2,\mu)$ and $\Gamma^{(n)}_R(p_i,g,m^2,\mu)$,
respectively, where $p_i$ $(i=1,\cdots,n)$ indicate momenta.
The energy scale $\mu$ indicates the renormalized point.
$\Gamma^{(n)}_R$ has the mass dimension $n+d-nd/2$:
$[\Gamma^{(n)}_R]=\mu^{n+d-nd/2}$.  These quantities are
related by the renormalization constant $Z_{\phi}$ as
\begin{equation}
\Gamma^{(n)}_R(p_i,g,m^2,\mu)= Z_{\phi}^{n/2}
\Gamma^{(n)}_B(p_i,g_0,m_0^2,\mu).
\end{equation}
Here we consider the massless case and omit the mass.
Because the bare quantity $\Gamma^{(n)}_B$ is independent of $\mu$,
we have
\begin{equation}
\frac{d}{d\mu}\Gamma^{(n)}_B = 0.
\end{equation}
This leads to
\begin{equation}
\mu\frac{d}{d\mu}\left( Z_{\phi}^{-n/2}\Gamma^{(n)}_R\right)=0.
\end{equation}
Then we obtain the equation for $\Gamma^{(n)}_R$:
\begin{equation}
\left( \mu\frac{\partial}{\partial\mu}+\mu\frac{\partial g}{\partial\mu}
\frac{\partial}{\partial g}-\frac{n}{2}\gamma_{\phi}\right)
\Gamma^{(n)}_R(p_i,g,\mu)=0,
\end{equation}
where $\gamma_{\phi}$ is defined as
\begin{equation}
\gamma_{\phi}= \mu\frac{\partial}{\partial\mu}\ln Z_{\phi}.
\end{equation}

A general solution of the renormalization group equation is
written as
\begin{equation}
\Gamma^{(n)}_R(p_i,g,\mu)= \exp\left( \frac{n}{2}
\int_{g_1}^{g}\frac{\gamma_{\phi}(g')}{\beta(g')}dg'\right)
f^{(n)}(p_i,g,\mu),
\end{equation}
where
\begin{equation}
f^{(n)}(p_i,g,\mu)= F\left(p_i,\ln\mu-\int_{g_1}^g
\frac{1}{\beta(g')}dg'\right),
\end{equation}
for a function $F$ and a constant $g_1$.
We suppose that $\beta(g)$ has a zero at $g=g_c$.
Near the fixed point $g_c$, by approximating $\gamma_{\phi}(g')$
by $\gamma_{\phi}(g_c)$, $\Gamma^{(n)}_R$ is expressed as
\begin{equation}
\Gamma^{(n)}_R(p_i,g_c,\mu)= \mu^{\frac{n}{2}\gamma_{\phi}(g_c)}
f^{(n)}(p_i,g_c,\mu).
\end{equation}
We define $\gamma(g)$ as
\begin{equation}
\gamma(g)\ln\mu= \int_{g_1}^g\frac{\gamma_{\phi}(g')}{\beta(g')}
dg',
\end{equation}
then we obtain
\begin{equation}
\Gamma^{(n)}_R(p_i,g,\mu)= \mu^{\frac{n}{2}\gamma(g)}
f^{(n)}(p_i,g,\mu).
\end{equation}
Under the scaling $p_i\rightarrow \rho p_i$,
$\Gamma^{(n)}_R$ is expected to behave as
\begin{equation}
\Gamma^{(n)}_R(\rho p_i,g_c,\mu)= \rho^{n+d-bd/2}
\Gamma^{(n)}_R(p_i,g_c,\mu/\rho),
\end{equation}
because $\Gamma^{(n)}_R$ has the mass dimension
$n+d-nd/2$.  In fact, the diagram in Fig.4(b) gives a
contribution being proportional to
\begin{eqnarray}
&&g^2(\mu^{4-d})^2\int d^dq\frac{1}{q^2(\rho p+q)^2} \nonumber\\
&=& g^2(\mu^{4-d})^2\rho^{d-4}\int d^dq\frac{1}{q^2(p+q)^2}
\nonumber\\
&=& \rho^{4-d}g^2\left(\frac{\mu}{\rho}\right)^{2(4-d)}
\int d^dq\frac{1}{q^2(p+q)^2},
\end{eqnarray}
after the scaling $p_i\rightarrow \rho p_i$ for $n=4$.
We employ eq.(58) for $n=2$:
\begin{eqnarray}
\Gamma^{(2)}(\rho p_i,g_c,\mu) &=& \rho^2
\Gamma^{(2)}_R(p_i,g_c,\mu/\rho) \nonumber\\
&=& \rho^2\left(\frac{\mu}{\rho}\right)^{\gamma}
f^{(2)}(p_i,g_c,\mu/\rho)\nonumber\\
&=& \rho^{2-\gamma}\mu^{\gamma}f^{(2)}(p_i,g_c,\mu/\rho)
\nonumber\\
&=& \rho^{2-\gamma}\Gamma^{(2)}_R(p_i,g_c,\mu/\rho).
\end{eqnarray}
This indicates
\begin{equation}
\Gamma^{(2)}(p)= p^{2-\eta}= p^{2-\gamma}
=(p^2)^{1-\gamma/2}.
\end{equation}
Thus the anomalous dimension $\eta$ is given by $\eta=\gamma$.
From the definition of $\gamma(g)$ in eq.(56), we have
\begin{equation}
\gamma_{\phi}(g)= \gamma(g)+\beta(g)\frac{\partial\gamma(g)}{\partial g}
\ln\mu.
\end{equation}
At the fixed point $g=g_c$, this leads to
\begin{equation}
\eta=\gamma=\gamma(g_c)=\gamma_{\phi}(g_c).
\end{equation}
The exponent $eta$ shows the fluctuation effect near the
critical point.

The Green function $G(p)=\Gamma^{(2)}(p)^{-1}$ is given by
\begin{equation}
G(p)= \frac{1}{p^{2-\eta}}.
\end{equation}
The Fourier transform of $G(p)$ in $d$ dimensions is evaluated as
\begin{eqnarray}
G({\bf r}) &=& \int \frac{1}{p^{2-\eta}}e^{ip\cdot r}d^d p
\nonumber\\
&=& \Omega_d\frac{1}{r^{d-2+\eta}}
\frac{\pi}{2\Gamma(4-\eta-d)\sin((4-\eta-d)\pi/2)}.
\nonumber\\
\end{eqnarray}
When $4-\eta-d$ is small near four dimensions, $G({\bf r})$
is approximated as
\begin{equation}
G(r) \simeq \Omega_d\frac{1}{r^{d-2+\eta}}.
\end{equation}

The definition of $\gamma_{\phi}$ in eq.(52) results in
\begin{equation}
\gamma_{\phi}(g)= \mu\frac{\partial g}{\partial\mu}
\frac{\partial}{\partial g}\ln Z_{\phi}
= \beta(g)\frac{\partial}{\partial g}\ln Z_{\phi}.
\end{equation}
Up to the lowest order of $g$, $\gamma_{\phi}$ is given by
\begin{eqnarray}
\gamma_{\phi}&=& \left( -\frac{1}{8\epsilon}\frac{N+1}{9}
\frac{1}{(8\pi^2)^2}g\right)\beta(g)+O(g^3)\nonumber\\
&=& \frac{N+2}{72}\frac{1}{(8\pi^2)^2}g^2+O(g^3).
\end{eqnarray}
At the critical point $g=g_c$, where
\begin{equation}
\frac{1}{8\pi^2}g_c = \frac{6\epsilon}{N+8},
\end{equation}
the anomalous dimension is given as
\begin{equation}
\eta= \gamma_{\phi}(g_c)= \frac{N+2}{2(N+8)^2}\epsilon^2
+O(\epsilon^3).
\end{equation}
For $N=1$ and $\epsilon=1$, we have $\eta=1/54$.

\begin{figure}[htbp]
\begin{center}
  \includegraphics[height=2.0cm]{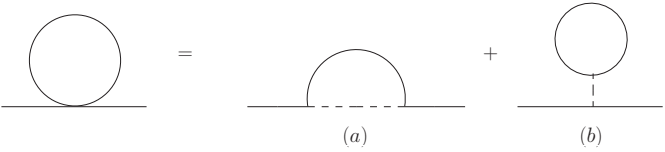}
\caption{
Corrections to the mass term.  Multiplicity weights are
8 for (a) and $2N$ for (b).
}
\label{fig6}       
\end{center}
\end{figure}

\subsection{Mass renormalization}

Let us consider the massive case with $m\neq 0$.
This corresponds to the case with $T>T_c$ in a phase transition.
The bare mass $m_0$ and renormalized mass $m$ are related 
through the relation $m^2=m_0^2Z_{\phi}/Z_2$.
The condition $\mu\partial m_0/\partial\mu=0$ leads to
\begin{equation}
\mu\frac{\partial\ln m^2}{\partial\mu}=\mu\frac{\partial}{\partial\mu}
\ln\left(\frac{Z_{\phi}}{Z_2}\right).
\end{equation}
From eq.(50), the equation for $\Gamma^{(n)}_R$ reads
\begin{eqnarray}
&&\Big[ \mu\frac{\partial}{\partial\mu}+\beta(g)\frac{\partial}{\partial g}
-\frac{n}{2}\gamma_{\phi}
+\mu\frac{\partial}{\partial\mu}\ln\left(\frac{Z_{\phi}}{Z_2}\right)
\cdot m^2\frac{\partial}{\partial m^2}\Big]\nonumber\\
&& ~~~~ \cdot \Gamma^{(n)}_R(p_i,g,\mu,m^2)=0.
\end{eqnarray}
We define the exponent $\nu$ by
\begin{equation}
\frac{1}{\nu}-2 = \mu\frac{\partial}{\partial\mu}
\ln\left( \frac{Z_{2}}{Z_{\phi}}\right),
\end{equation}
then
\begin{eqnarray}
&&\Big[ \mu\frac{\partial}{\partial\mu}+\beta(g)\frac{\partial}{\partial g}
-\frac{n}{2}\gamma_{\phi}
-\left( \frac{1}{\nu}-2\right)
m^2\frac{\partial}{\partial m^2}\Big]\nonumber\\
&& ~~~~ \cdot \Gamma^{(n)}_R(p_i,g,\mu,m^2)=0.
\end{eqnarray}
At the critical point $g=g_c$, we obtain
\begin{equation}
\Big[ \mu\frac{\partial}{\partial\mu}
-\frac{n}{2}\eta-\zeta m^2\frac{\partial}{\partial m^2}\Big]
\Gamma^{(n)}_R(p_i,g_c,\mu,m^2)=0,
\end{equation}
where $\gamma_{\phi}=\eta$ and we set
\begin{equation}
\zeta= \frac{1}{\nu}-2.
\end{equation}
At $g=g_c$, $\Gamma^{(n)}_R$ has the form
\begin{equation}
\Gamma^{(n)}_R(p_i,g_c,\mu,m^2)= \mu^{n/2}F^{(n)}(p_i,\mu m^{2/\zeta}),
\end{equation}
because this satisfies eq.(75).

In the scaling $p_i\rightarrow \rho p_i$, we adopt
\begin{equation}
\Gamma^{(n)}_R(\rho p_i,g_c,\mu,m^2)= \rho^{n+d-nd/2}
\Gamma^{(n)}_R(p_i,g_c,\mu/\rho,m^2/\rho^2).
\end{equation}
From eq.(77), we have
\begin{eqnarray}
\Gamma^{(n)}_R(k_i,g_c,\mu,m^2)&=& \rho^{n+d-nd/2-n\eta/2}
\mu^{n\eta/2} \nonumber\\
&& \cdot F^{(n)}(\rho^{-1}k_i,\rho^{-1}\mu(\rho^{-2}m^2)^{1/\zeta}),
\nonumber\\
\end{eqnarray}
where we put $\rho p_i=k_i$.
We assume that $F^{(n)}$ depends only on $\rho^{-1}k_i$.
We choose $\rho$ as
\begin{equation}
\rho= (\mu m^{2/\zeta})^{\zeta/(\zeta+2)}
=\mu\left(\frac{m^2}{\mu^2}\right)^{1/(\zeta+2)}.
\end{equation}
This satisfies $\rho^{-1}\mu(\rho^{-2}m^2)^{1/\zeta}=1$ and
results in
\begin{eqnarray}
&&\Gamma^{(n)}_R(k_i,g_c,\mu,m^2) \nonumber\\
&=& \mu^{d+\frac{n}{2}(2-d-\eta)}
\cdot \left(\frac{m^2}{\mu^2}\right)^{\left(d+\frac{n}{2}(2-d-\eta)\right)
\frac{1}{\zeta+2}}\nonumber\\
&& \cdot \mu^{\frac{n}{2}\eta} F^{(n)}\left(\mu^{-1}
\left(\frac{m^2}{\mu^2}\right)^{-\frac{1}{\zeta+2}}k_i\right).
\end{eqnarray}
We take $\mu$ as a unit by setting $\mu=1$, so that $\Gamma^{(n)}_R$ is
written as
\begin{equation}
\Gamma^{(n)}_R(k_i,g_c,1,m^2)= m^{2\nu \{d+\frac{n}{2}(2-d-\eta)\}}F^{(n)}(k_im^{-2\nu}),
\end{equation}
because $\zeta+2=1/\nu$.  We define the correlation length $\xi$ by
\begin{equation}
(m^2)^{-\nu}=\xi.
\end{equation}
The two-point function for $n=2$ is written as
\begin{equation}
\Gamma^{(2)}_R(k,m^2)= m^{2\nu(2-\eta)}F^{(2)}(km^{-2\nu}).
\end{equation}

Now let us turn to the evaluation of $\nu$.  Since 
$\gamma_{\phi}=\mu\partial\ln Z_{\phi}/\partial\mu$, from eq.(73) $\nu$ is
given by
\begin{equation}
\frac{1}{\nu}=2+\mu\frac{\partial}{\partial\mu}\ln\left(\frac{Z_2}{Z_{\phi}}\right)
= 2+\beta(g)\frac{\partial}{\partial g}\ln Z_2-\gamma_{\phi}(g).
\end{equation}
The renormalization constant $Z_2$ is determined from the corrections to
the bare mass $m_0$.  The one-loop correction, shown in Fig.6, is given by
\begin{equation}
\Sigma(p^2)= i\frac{N+2}{6}g\int\frac{d^dk}{(2\pi)^d}\frac{1}{k^2-m_0^2},
\end{equation} 
where the multiplicity factor is $(8+4N)/4!$.  This is regularized as
\begin{equation}
\Sigma(p^2)= \frac{N+2}{6}g\int\frac{d^dk}{(2\pi)^d}\frac{1}{k_E^2+m_0^2}
= -\frac{N+2}{6}g\frac{1}{8\pi^2}m_o^2\frac{1}{\epsilon},
\end{equation}
for $d=4-\epsilon$.  Therefore the renormalized mass is
\begin{equation}
m^2=m_0^2+\Sigma(p^2)= m_0^2\left( 1-\frac{N+2}{6\epsilon}\frac{1}{8\pi^2}g\right).
\end{equation}
$Z_2$ is determined to cancel the divergence in the form $m^2Z_2/Z_{\phi}$.
The result is
\begin{equation}
Z_2= 1+\frac{N+2}{6\epsilon}\frac{1}{8\pi^2}g.
\end{equation}
Then, we have
\begin{equation}
\beta(g)\frac{\partial}{\partial g}\ln Z_2= -\frac{N+2}{6}\frac{1}{8\pi^2}g
+O(g^2).
\end{equation}
The eq.(85) is written as
\begin{equation}
\frac{1}{\nu}=2-\frac{N+2}{6}\frac{1}{8\pi^2}g_c-\eta
=2-\frac{N+2}{N+8}\epsilon+O(\epsilon^2),
\end{equation}
where we put $g=g_c$ and used $\eta=\gamma_{\phi}=(N+2)/(2(N+8)^2)\cdot\epsilon$.
Now the exponent $\nu$ is
\begin{equation}
\nu= \frac{1}{2}\left( 1+\frac{N+2}{2(N+8)}\epsilon\right)+O(\epsilon^2).
\end{equation}
In the mean-field approximation we have $\nu=1/2$.
This formula of $\nu$ contains the fluctuation effect near the critical
point.  For $N=1$ and $\epsilon=1$, we have
$\nu=1/2+1/12=7/12$.

\section{Non-linear Sigma Model}
\subsection{Lagrangian}

The Lagrangian of the non-linear sigma model is
\begin{equation}
\mathcal{L}= \frac{1}{2g}(\partial_{\mu}\phi)^2,
\end{equation}
where $\phi$ is a real $N$-component $\phi=(\phi_1,\cdots,\phi_N)$
with the constraint $\phi^2=1$.
This model has an $O(N)$ invariance.  THe field $\phi$ is represented as
\begin{equation}
\phi= (\sigma,\pi_1,\pi_2,\cdots,\pi_{N-1}),
\end{equation}
with the condition $\sigma^2+\pi_1^2+\cdots+\pi_{N-1}^2=1$.
The fields $\pi_i$ ($i=1,\cdots,N-1$) represent fluctuations.
The Lagrangian is given by
\begin{equation}
\mathcal{L}= \frac{1}{2g}\left( (\partial_{\mu}\sigma)^2
+(\partial_{\mu}\pi_i)^2\right),
\end{equation}
where summation is assumed for index $i$.  In this Section we consider
the Euclidean Lagrangian from the beginning.
Using the constraint $\sigma^2+\pi_i^2=1$, the Lagrangian is written
in the form:
\begin{eqnarray}
\mathcal{L}&=& \frac{1}{2g}(\partial_{\mu}\pi_i)^2
+\frac{1}{2g}\frac{1}{1-\pi_i^2}(\pi_i\partial_{\mu}\pi_i)^2
\\
&=& \frac{1}{2g}(\partial_{\mu}\pi_i)^2+\frac{1}{2g}
(\pi_i\partial_{\mu}\pi_i)^2+\cdots.
\end{eqnarray}
The second term in the right-hand side indicates the interaction
between $\pi_i$ fields.  The diagram for this interaction is
shown in Fig.7.

\begin{figure}[htbp]
\begin{center}
  \includegraphics[height=2.5cm]{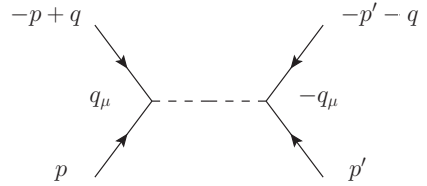}
\caption{
Lowest-order interaction for $\pi_i$.
}
\label{fig7}       
\end{center}
\end{figure}

Here let us check the dimension of the field and coupling constant.
Since $[\mathcal{L}]=\mu^d$, we obtain $[\pi]=\mu^0$ (dimensionless)
and $[g]=\mu^{2-d}$.  $g_0$ and $g$ are used to denote the bare
coupling constant and renormalized coupling constant, respectively.
The bare and renormalized fields are indicated by $\pi_{Bi}$
and $\pi_{Ri}$, respectively.
We define the renormalization constants $Z_g$ and $Z$ by
\begin{eqnarray}
g_0 &=& g\mu^{2-d}Z_g, \\
\pi_{Bi} &=& \sqrt{Z}\pi_{Ri},
\end{eqnarray}
where $g$ is the dimensionless coupling constant.  Then,
the Lagrangian is expressed in terms of renormalized quantities:
\begin{equation}
\mathcal{L}= \frac{\mu^{d-2}Z}{2gZ_g}\left( (\partial_{\mu}\pi_{Ri})^2
+\frac{1}{4}(\partial_{\mu}\pi^2_{Ri})^2+\cdots \right).
\end{equation}
In order to avoid the infrared divergence at $d=2$, we add the
Zeeman term to the Lagrangian which is written as
\begin{eqnarray}
\mathcal{L}_Z &=& \frac{H_B}{g_0}\sigma
= \frac{H_B}{g_0}\left( 1-\frac{Z}{2}\pi_{Ri}^2
-\frac{Z^2}{8}\pi_{Ri}^4+\cdots\right)\\
&=& {\rm const.}-H_B\frac{Z}{2gZ_g}\mu^{d-2}\pi_{Ri}^2
-H_B\frac{Z^2}{8gZ_g}\mu^{d-2}(\pi_{Ri}^2)^2.\nonumber\\
\end{eqnarray}
Here $H_B$ is the bare magnetic field and the renormalized
magnetic field $H$ is defined as
\begin{equation}
H= \frac{\sqrt{Z}}{Z_g}H_B.
\end{equation}
Then, the Zeeman term is given by
\begin{equation}
\mathcal{L}_Z = {\rm const.}-\frac{\sqrt{Z}}{2g}H\mu^{d-2}\pi_{Ri}^2
-\frac{Z^{3/2}}{8g}H\mu^{d-2}(\pi_{Ri}^2)^2+\cdots .
\end{equation}

\subsection{Two-point function}

The diagrams for the two-point function $\Gamma^{(2)}(p)=G^{(2)}(p)^{-1}$
are shown in Fig.8.  The contributions in Fig. 8(c) and 8(d) come 
from the magnetic field.  The term (b) in Fig. 8 gives
\begin{equation}
I_b= \int\frac{d^dk}{(2\pi)^d}\frac{(k+p)^2}{k^2+H}
=(p^2-H)\int\frac{d^dk}{(2\pi)^d}\frac{1}{k^2+H},
\end{equation}
where we used the formula in the dimensional regularization given as
\begin{equation}
\int d^dk=0.
\end{equation}
Near two dimensions, $d=2+\epsilon$, the integral is regularized as
\begin{eqnarray}
I_b&=& (p^2-H)\frac{\Omega_d}{(2\pi)^d}H^{\frac{d}{2}-1}
\Gamma\left(\frac{d}{2}\right)\Gamma\left(1-\frac{d}{2}\right)
\nonumber\\
&=& -(p^2-H)\frac{\Omega_d}{(2\pi)^d}\frac{1}{\epsilon}.
\end{eqnarray}
The term $I_c$ in Fig. 8(c) just cancels with $-H$ in $I_b$.
The contribution $I_d$ in Fig. 8(d) has the multiplicity
$2\times 2\times (N-1)$ because $(\pi_i)$ has $N-1$ components.
$I_d$ is evaluated as
\begin{equation}
I_d= \frac{1}{8}\cdot 4(N-1)\int\frac{d^dk}{(2\pi)^d}
\frac{1}{k^2+H}= -\frac{\Omega_d}{(2\pi)^d}\frac{N-1}{2}
\frac{1}{\epsilon}.
\end{equation}
As a result, up to the one-loop order the two-point function is
\begin{equation}
\Gamma^{(2)}(p)= \frac{Z}{Z_gg}p^2+\frac{\sqrt{Z}}{g}H
-\frac{1}{\epsilon}\left( p^2+\frac{N-1}{2}H\right),
\end{equation}
where the factor $\Omega_d/(2\pi)^d$ is included in $g$ for
simplicity.  To remove the divergence, we choose
\begin{eqnarray}
\frac{Z}{Z_g} &=& 1+\frac{g}{\epsilon},\\
\sqrt{Z} &=& 1+\frac{N-1}{2\epsilon}g.
\end{eqnarray}
This set of equations gives
\begin{eqnarray}
Z_g &=& 1+\frac{N-2}{\epsilon}g+O(g^2),\\
Z &=& 1+\frac{N-1}{\epsilon}g+O(g^2).
\end{eqnarray}

The case $N=2$ is a special case where we have $Z_g=1$.
This will hold even when including higher order corrections.
For $N=2$, we have one $\pi$ field satisfying
\begin{equation}
\sigma^2+\pi^2=1.
\end{equation}
When we parameterize $\sigma$ and $\pi$ as $\sigma=\cos\theta$ and
$\pi=\sin\theta$, the Lagrangian is
\begin{equation}
\mathcal{L}= \frac{1}{2g}\left( (\partial_{\mu}\sigma)^2
+(\partial_{\mu}\pi)^2 \right)
=\frac{1}{2g}(\partial_{\mu}\theta)^2.
\end{equation}
If we disregard the region of $\theta$, $0\le\theta \le 2\pi$,
the field $\theta$ is a free field suggesting that $Z_g=1$.

\begin{figure}[htbp]
\begin{center}
  \includegraphics[height=1.7cm]{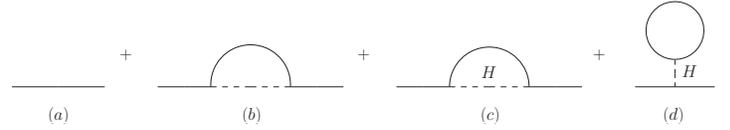}
\caption{
Diagrams for the two-point function.  The diagrams (c) and (d)
come from the Zeeman term.
}
\label{fig8}       
\end{center}
\end{figure}

\subsection{Renormalization group equations}

The beta function $\beta(g)$ of the coupling constant $g$ is defined by
\begin{equation}
\beta(g)= \mu\frac{\partial g}{\partial\mu},
\end{equation}
where the bare quantities are fixed in calculating the derivative.
Since $\mu\partial g_0/\partial\mu=0$, the beta function is derived as
\begin{equation}
\beta(g) = \frac{\epsilon g}{1+g\frac{\partial}{\partial g}\ln Z_g}
= \epsilon g-(N-2)g^2+0(g^3),
\end{equation}
for $d=2+\epsilon$.
The beta function is shown in Fig.9 as a function of $g$.
We mention here that the coefficient $N-2$ of $g^2$ term is
related with the Casimir invariant of the symmetry group
$O(N)$\cite{bre80,yan16b}.

\begin{figure}[ht]
\includegraphics[width=5.8cm]{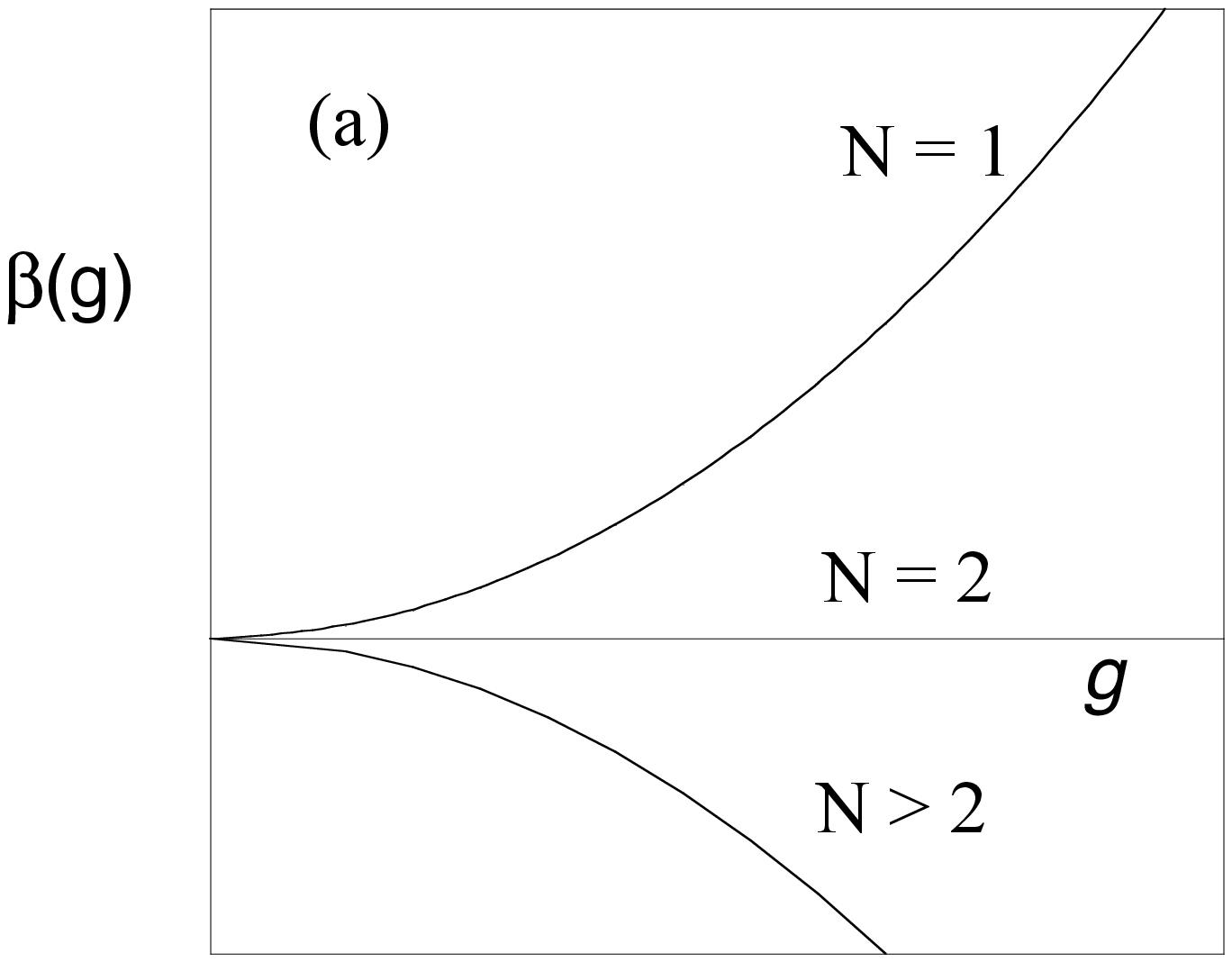}
\includegraphics[width=5.8cm]{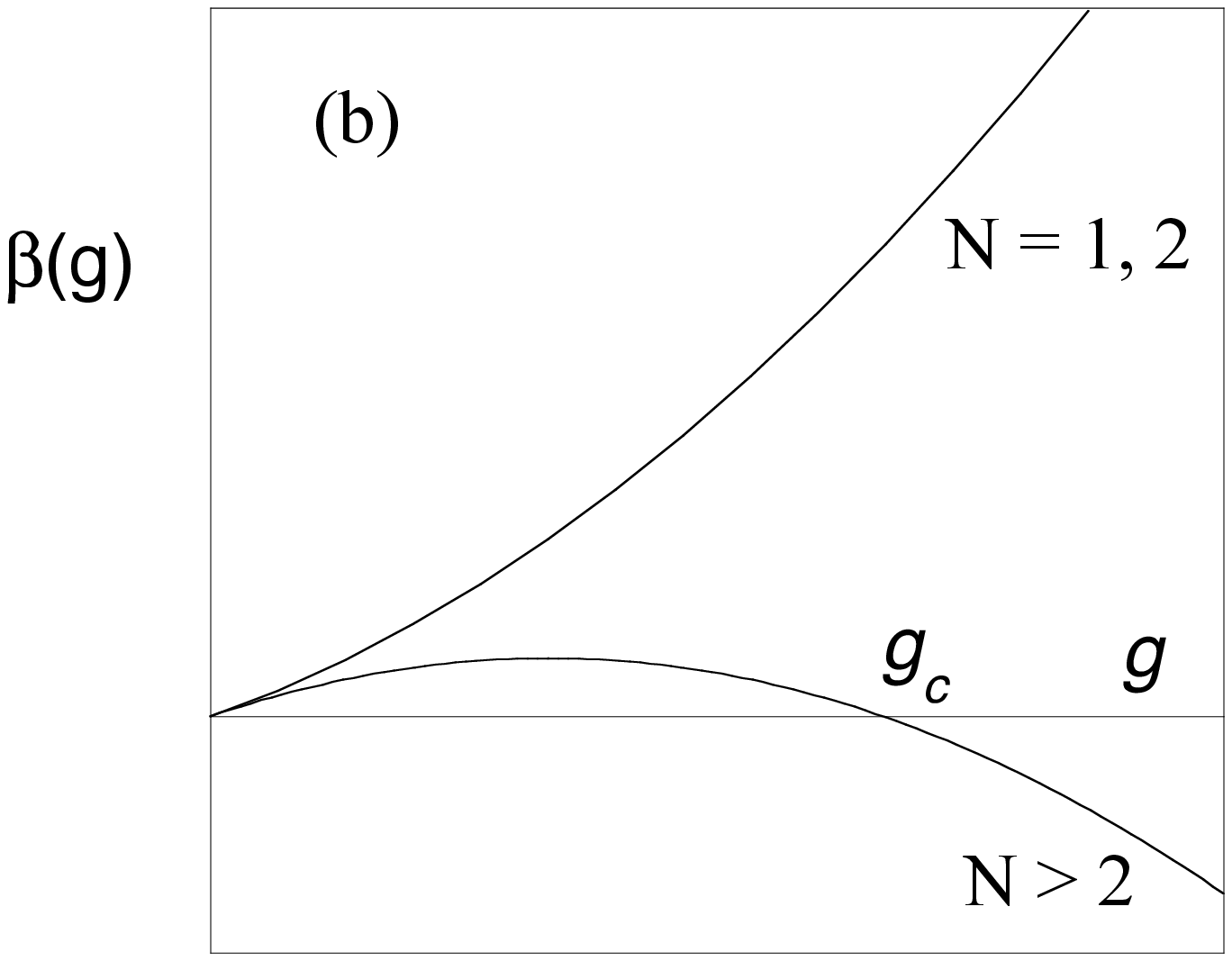}
 \caption{
The beta function $\beta(g)$ as a function of $g$ for $d=2$ (a) and
$d>2$ (b).  There is a fixed point for $N>2$ and $d>2$.
$\beta(g)$ is negative for $d>2$ and $N>2$ which indicates that
the model exhibits an asymptotic freedom.
 }
\end{figure}

In the case of $N=2$ and $d=2$, $\beta(g)$ vanishes.
This case corresponds to the classical XY model as mentioned above and
there may be a Kosterlitz-Thouless transition.
The Kosterlitz-Thouless transition point cannot be obtained by a
perturbative expansion in $g$.

In two dimensions $d=2$, $\beta(g)$ shows asymptotic freedom for
$N>2$.  The coupling constant $g$ approaches zero in high-energy
limit $\mu\rightarrow\infty$ in a similar way to QCD.
For $N=1$, $g$ increases as $\mu\rightarrow\infty$ as in the case of
QED.  When $d>2$, there is a fixed point $g_c$:
\begin{equation}
g_c=\frac{\epsilon}{N-2},
\end{equation}
for $N>2$.  There is a phase transition for $N>2$ and $d>2$.

Let us consider the n-point function $\Gamma^{(n)}(k_i,g,\mu,H)$.
The bare and renormalized n-point functions are introduced similarly
and they are related by the renormalization constant $Z$:
\begin{equation}
\Gamma_R^{(n)}(k_i,g,\mu,H)= Z^{n/2}\Gamma_B^{(n)}(k_i,g_0,\mu,H_B).
\end{equation}
From the condition that the bare function $\Gamma_B^{(m)}$ is
independent of $\mu$, $\mu d\Gamma_B^{(n)}/d\mu=0$, the
renormalization group equation is followed:
\begin{eqnarray}
&&\Big[ \mu\frac{\partial}{\partial\mu}
+\mu\frac{\partial g}{\partial\mu}\frac{\partial}{\partial g}
-\frac{n}{2}\zeta(g)+\left( \frac{1}{2}\zeta(g)+\frac{1}{g}\beta(g)
-(d-2)\right)\nonumber\\
&& \cdot H\frac{\partial}{\partial H}\Big] 
 \Gamma_R^{(n)}(k_i,g,\mu,H)=0,
\end{eqnarray}
where we defined
\begin{equation}
\zeta(g)= \mu\frac{\partial}{\partial\mu}\ln Z= \beta(g)
\frac{\partial}{\partial g}\ln Z.
\end{equation}
We used the relation between renormalized magnetic field $H$ and bare
field $H_B$ in eq. (103) in deriving eq. (120).  
From eq. (113), $\zeta(g)$ is given by
\begin{equation}
\zeta(g)= (N-1)g+O(g^2).
\end{equation}

Let us define the correlation length $\xi=xi(g,\mu)$.
Because the correlation length near the transition point will not
depend on the energy scale, it should satisfy
\begin{equation}
\mu\frac{d}{d\mu}\xi(g,\mu)= \left( \mu\frac{\partial}{\partial\mu}
+\beta(g)\frac{\partial}{\partial g}\right)\xi(g,\mu)=0.
\end{equation}
We adopt the form $\xi=\mu^{-1}f(g)$ for a function $f(g)$, so that
we have
\begin{equation}
\beta(g)\frac{df(g)}{dg}=f(g).
\end{equation}
This indicates
\begin{equation}
f(g)= C\exp\left( \int_{g_*}^g\frac{1}{\beta(g')}dg'\right),
\end{equation}
where $C$ and $g_*$ are constants.
In two dimensions ($\epsilon=0$), the beta function in eq. (117) gives
\begin{equation}
\xi= C\mu^{-1}\exp\left( \frac{1}{N-2}\left(\frac{1}{g}-\frac{1}{g_*}\right)
\right).
\end{equation}
When $N>2$, $\xi$ diverges as $g\rightarrow 0$, namely, the mass
being proportional to $\xi^{-1}$ vanishes in this limit.
When $d>2$ ($\epsilon>0$), there is a finite-fixed point $g_c$.
We approximate $\beta(g)$ near $g=g_c$ as
\begin{equation}
\beta(g)\approx a(g-g_c),
\end{equation}
with $a<0$, $\xi$ is
\begin{equation}
\xi= \mu^{-1}\exp\left( \frac{1}{a}\ln\Big|\frac{g-g_c}{g_*-g_c}\Big|\right).
\end{equation}
Near the critical point $g\approx g_c$, $\xi$ is approximated as
\begin{equation}
\xi^{-1}\simeq \mu|g-g_c|^{1/|a|}.
\end{equation}
This means that $\xi\rightarrow \infty$ as $g\rightarrow g_c$.
We define the exponent $\nu$ by
\begin{equation}
\xi^{-1}\simeq |g-g_c|^{\nu},
\end{equation}
then we have
\begin{equation}
\nu= -\frac{1}{\beta'(g_c)}.
\end{equation}
Since $\beta'(g_c)=\epsilon-2(N-2)g_c=-\epsilon$, this gives
\begin{equation}
\frac{1}{\nu}= \epsilon+O(\epsilon^2)= d-2+O(\epsilon^2).
\end{equation}
Including the higher order terms, $\nu$ is given as
\begin{equation}
\frac{1}{\nu}=d-2+\frac{(d-2)^2}{N-2}+\frac{(d-2)^3}{2(N-2)}
+O(\epsilon^4).
\end{equation}

\subsection{2D Quantum Gravity}

A similar renormalization group equation is derived for the two-dimensional
quantum gravity.  The space structure is written by the metric tensor
$g_{\mu\nu}$ and the curvature $R$.
The quantum gravity Lagrangian is
\begin{equation}
\mathcal{L}= -\frac{1}{16\pi G}\sqrt{g}R,
\end{equation}
where $g$ is the determinant of the matrix $(g_{\mu\nu})$ and $G$ is
the coupling constant.  The beta function for $G$ was calculated
as\cite{gas78,chr78,smo82,kaw90}.
\begin{equation}
\beta(G)= \epsilon G-bG^2,
\end{equation}
for $d=2+\epsilon$ with a constant $b$.
This has the same structure as that for the non-linear sigma model.

\section{Sine-Gordon Model}

\subsection{Lagrangian}

The two-dimensional sine-Gordon model has attracted a lot of
attention\cite{col75,bre79,das79,zam79,raj82,man04,yan16b,man75,jos77,sch77,sam78,
wie78,kog79,ami80,hua91,nan04,nag09}.
The Lagrangian of the sine-Gordon model is given by
\begin{equation}
\mathcal{L}= \frac{1}{2t_0}(\partial_{\mu}\phi)^2
+\frac{\alpha_0}{t_0}\cos\phi,
\end{equation}
where $\phi$ is a real scalar field, and $t_0$ and $\alpha_0$ are
bare coupling constants.
We also use the Euclidean notation in this Section.
The second term is the potential energy of the scalar field.
We adopt that $t$ and $\alpha$ are positive.  The renormalized coupling
constants are denoted as $t$ and $\alpha$, respectively.
The dimensions of $t$ and $\alpha$ are $[t]=\mu^{2-d}$ and
$[\alpha]=\mu^2$.  The scalar field $\phi$ is dimensionless in this
representation.  The renormalization constants $Z_t$ and $Z_{\alpha}$
are defined as follows:
\begin{equation}
t_0= t\mu^{2-d}Z_t,~~\alpha_0= \alpha\mu^2 Z_{\alpha}.
\end{equation}
Here, the energy scale $\mu$ is introduced so that $t$ and $\alpha$
are dimensionless.  The Lagrangian is written as
\begin{equation}
\mathcal{L}= \frac{\mu^{d-2}}{2tZ_t}(\partial_{\mu}\phi)^2
+\frac{\mu^d\alpha Z_{\alpha}}{tZ_t}\cos\phi.
\end{equation}
We introduce the renormalized field $\phi_R$ by
$\phi_B=\sqrt{Z_{\phi}}\phi_R$ where $Z_{\phi}$ is the renormalization
constant.  Then the Lagrangian is
\begin{equation}
\mathcal{L}= \frac{\mu^{d-2}Z_{\phi}}{2tZ_t}(\partial_{\mu}\phi)^2
+\frac{\mu^d\alpha Z_{\alpha}}{tZ_t}\cos(\sqrt{Z_{\phi}}\phi),
\end{equation}
where $\phi$ stands for the renormalized field $\phi_R$.

\subsection{Renormalization of $\alpha$}

We investigate the renormalization group procedure for the sine-Gordon
model on the basis of the dimensional regularization method.
Fitst, let us consider the renormalization of the potential term.
The lowest-order contributions are given by tadpole diagrams.
We use the expansion $\cos\phi=1-\frac{1}{2}\phi^2+\frac{1}{4!}\phi^4+\cdots$.
Then the corrections to the cosine term are evaluated as follows.
The constant term is renormalized as
\begin{eqnarray}
&& 1-\frac{1}{2}\langle\phi^2\rangle+\frac{1}{4!}\langle\phi^4\rangle-\cdots
= 1-\frac{1}{2}\langle\phi^2\rangle+\frac{1}{2}\left(\frac{1}{2}\langle\phi^2\rangle
\right)^2 -\cdots \nonumber\\
&& ~~~~ = \exp\left( -\frac{1}{2}\langle\phi^2\rangle\right).
\end{eqnarray}
Similarly, the $\phi^2$ is renormalized as
\begin{eqnarray}
&&-\frac{1}{2}\phi^2+\frac{1}{4!}6\langle\phi^2\rangle\phi^2
-\frac{1}{6!}15\cdot 3\langle\phi^2\rangle^2\phi^2+\cdots
\nonumber\\
&&~~~~ = \exp\left(-\frac{1}{2}\langle\phi^2\rangle\right)
\left(-\frac{1}{2}\phi^2\right).
\end{eqnarray}
Hence the $\alpha Z_{\alpha}\cos\left(\sqrt{Z_{\phi}}\phi\right)$
is renormalized to
\begin{eqnarray}
&&\alpha Z_{\alpha}\exp\left( -\frac{1}{2}Z_{\phi}\langle\phi^2\rangle\right)
\cos(\sqrt{Z_{\phi}}\phi) \nonumber\\
&& ~~~ \simeq \alpha Z_{\alpha}\left(1-\frac{1}{2}
Z_{\phi}\langle\phi^2\rangle+\cdots\right) 
\cos\left(\sqrt{Z_{\phi}}\phi\right).
\end{eqnarray}
The expectation value $\langle\phi^2\rangle$ is regularized as
\begin{equation}
Z_{\phi}\langle\phi^2\rangle= t\mu^{2-d}Z_t\int\frac{d^dk}{(2\pi)^d}
\frac{1}{k^2+m_0^2}= -\frac{t}{\epsilon}\frac{\Omega_d}{(2\pi)^d},
\end{equation}
where $d=2+\epsilon$ and we included a mass $m_0$ to avoid the
infrared divergence.  The constant $Z_{\alpha}$ is determined to
cancel the divergence:
\begin{equation}
Z_{\alpha}= 1-\frac{t}{2}\frac{1}{\epsilon}\frac{\Omega_d}{(2\pi)^d}.
\end{equation}
From the equations $\mu\partial t_0/\partial\mu=0$ and
$\mu\partial\alpha_0/\partial\mu=0$, we obtain
\begin{eqnarray}
\mu\frac{\partial t}{\partial\mu}&=& (d-2)t-t\mu\frac{\partial\ln Z_t}{\partial\mu},\\
\mu\frac{\partial\alpha}{\partial\mu}&=& -2\alpha
-\alpha\mu\frac{\partial\ln Z_t}{\partial\mu}.
\end{eqnarray}
The beta function for $\alpha$ reads
\begin{equation}
\beta(\alpha)\equiv \mu\frac{\partial\alpha}{\partial\mu}
=-2\alpha+t\alpha\frac{1}{2}\frac{\Omega_d}{(2\pi)^d},
\end{equation}
where we set $\mu\partial t/\partial\mu=(d-2)t$ with $Z_t=1$ up to
the lowest order of $\alpha$.
The function $\beta(\alpha)$ has a zero at $t=t_c=8\pi$.

\subsection{Renormalization of the two-point function}

Let us turn to the renormalization of the coupling constant $t$.  
The renormalization of $t$ comes from the correction to $p^2$ term.
The lowest-order two-point function is
\begin{equation}
\Gamma_B^{(2)(0)}(p)= \frac{1}{t_0}p^2
=\frac{1}{t\mu^{2-d}Z_t}p^2.
\end{equation}
The diagrams that contribute to the two-point function are shown in
Fig.10\cite{ami80}.
These diagrams are obtained by expanding the cosine function as
$\cos\phi=1-(1/2)\phi^2+\cdots$.  First, we consider the Green function 
\begin{eqnarray}
G_0(x)&\equiv& Z_{\phi}\langle\phi(x)\phi(0)\rangle=t\mu^{2-d}Z_t
\int\frac{d^dp}{(2\pi)^d}\frac{e^{ip\cdot x}}{p^2+m_0^2} \nonumber\\
&=& t\mu^{2-d}Z_t\frac{\Omega_d}{(2\pi)^d}K_0(m_0|x|),
\end{eqnarray}
where $K_0$ is the 0-th modified Bessel function and $m_0$ is 
introduced to avoid the infrared singularity.  Because $\sinh I-I=I^3/3!+\cdots$,
the diagrams in Fig.10 are summed up to give 
\begin{equation}
\Sigma (p)= \int d^dx\big[ e^{ip\cdot x}(\sinh I-I)-(\cosh I-1)\big],
\end{equation}
where $I=G_0(x)$.  Since $\sinh I-I\simeq e^I/2$ and $\cosh I\simeq e^I/2$,
the diagrams in Fig.10 leads to
\begin{equation}
\Gamma_B^{(2)c}(p)= \frac{1}{2}\left(\frac{\alpha\mu^dZ_{\alpha}}{tZ_t}\right)^2
\int d^dx(e^{ip\cdot x}-1)e^{G_0(x)}.
\end{equation}
We use the expansion $e^{ip\cdot x}=1+ip\cdot x-(1/2)(p\cdot x)^2+\cdots$,
and keep the $p^2$ term.  We write deviation of $t$ from the fixed 
point $t_c=8\pi$ as $v$:
\begin{equation}
\frac{t}{8\pi}= 1+v,
\end{equation}
for $d=2$.  Using the asymptotic formula $K_0(x)\sim -\gamma-\ln(x/2)$ 
for small $x>0$, we obtain 
\begin{eqnarray}
\Gamma_B^{(2)c}(p)&=& \frac{1}{8}\left(\frac{\alpha\mu^d}{tZ_t}\right)^2
p^2(c_0m_0^2)^{-2-2v}\Omega_d\int_0^{\infty}dxx^{d+1}\nonumber\\
&& ~~~ \cdot \frac{1}{(x^2+a^2)^{2+2v}}
\nonumber\\
&=& -\frac{1}{8}p^2\left(\frac{\alpha\mu^d}{tZ_t}\right)^2(c_0M_0^2)^{-2}
\Omega_d\frac{1}{\epsilon}+O(v) \nonumber\\
&=& -\frac{1}{t\mu^{2-d}Z_t}p^2\frac{1}{32}\alpha^2\mu^{d+2}
(c_0m_0^2)^{-2}\frac{1}{\epsilon}+O(v),
\nonumber\\
\end{eqnarray}
where $c_0$ is a constant and $a=1/\mu$ is a small cutoff.
The divergence of $\alpha$ was absorbed by $Z_{\alpha}$.  Now the two-point
function up to this order is
\begin{equation}
\Gamma_B^{(2)}(p)= \frac{1}{t\mu^{2-d}Z_t}\Big[ p^2
-p^2\frac{1}{32}\alpha^2\mu^{d+2}(c_0m_0^2)^{-2}\frac{1}{\epsilon}\Big].
\end{equation}
The renormalized two-point function is $\Gamma_R^{(2)}=Z_{\phi}\Gamma_B^{(2)}$.
This indicates that 
\begin{equation}
\frac{Z_{\phi}}{Z_t}= 1+\frac{1}{32}\alpha^2\mu^{d+2}(c_0m_0^2)^{-2}
\frac{1}{\epsilon}.
\end{equation}
Then we can choose $Z_{\phi}=1$ and
\begin{equation}
Z_t= 1-\frac{1}{32}\alpha^2\mu^{d+2}(c_0m_0^2)^{-2}\frac{1}{\epsilon}.
\end{equation}
$Z_t/Z_{\phi}$ can be regarded as the renormalization constant of $t$ up 
to the order of $\alpha^2$, and thus we do not need the renormalization
constant $Z_{\phi}$ of the field $\phi$.  This means that we can 
adopt the bare coupling constants as $t_0=t\mu^{2-d}\bar{Z}_t$ with
$\bar{Z}_t=Z_t/Z_{\phi}$.

The renormalization function of $t$ is obtained from the equation
$\mu\partial t_0/\partial\mu=0$ for $t_0=t\mu^{2-d}Z_t$:
\begin{eqnarray}
\beta(t)&\equiv& \mu\frac{\partial t}{\partial\mu}=(d-2)t
+\frac{1}{32}(c_0 m_0^2)^{-2}\frac{t}{\epsilon} \nonumber\\
&& \cdot \left(2\alpha\mu^{d+2}\mu\frac{\partial\alpha}{\partial\mu} 
+(d+2)\alpha^2\mu^{d+2}\right) \nonumber\\
&=& (d-2)t+\frac{1}{32}t\alpha^2\mu^{d+2}(c_0 m_0^2)^{-2}+O(t^2).
\nonumber\\
\end{eqnarray}
Because the finite part of $G_0(x\rightarrow 0)$ is given by
$G_0(x\rightarrow 0)=-(1/2\pi)\ln(e^{\gamma}m_0/2\mu)$, we perform
the finite renormalization of $\alpha$ as
$\alpha\rightarrow \alpha c_0m_0^2a^2=\alpha c_0m_0^2\mu^{-2}$.
This results in
\begin{equation}
\beta(t)= (d-2)t+\frac{1}{32}t\alpha^2.
\end{equation}
As a result, we obtain a set of renormalization group equations for 
the sine-Gordon model:
\begin{eqnarray}
\beta(\alpha)&=& \mu\frac{\partial\alpha}{\partial\mu}=
-\alpha\left( 2-\frac{1}{4\pi}t\right), \\
\beta(t)&=& \mu\frac{\partial t}{\partial\mu}= (d-2)t
+\frac{1}{32}t\alpha^2.
\end{eqnarray}
Since the equation for $\alpha$ is homogeneous in $\alpha$, we can 
change the scale of $\alpha$ arbitrarily.  Thus, the numerical
coefficient of $t\alpha^2$ is not important.

\vspace{0.5cm}
\begin{figure}[ht]
\includegraphics[width=4.0cm,angle=90]{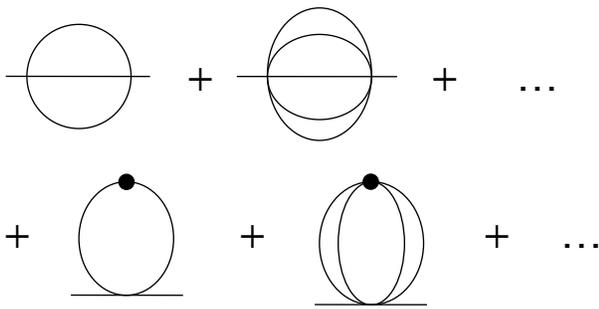}
 \caption{
Diagrams that contribute to the two-point function.
 }
\end{figure}

\subsection{Renormalization group flow}

Let us investigate the renormalization group flow in two dimensions.
This set of equations reduces to that of the Kosterlitz-Thouless
(K-T) transition.  We write $t=8\pi(1+v)$, and set $x=2v$ and
$y=\alpha/4$.  Then, the equations are
\begin{eqnarray}
\mu\frac{\partial x}{\partial\mu}&=& y^2, \\
\mu\frac{\partial y}{\partial\mu}&=& xy.
\end{eqnarray}
These are equations of K-T transition.  From these, we have
\begin{equation}
x^2-y^2= {\rm const}.
\end{equation}
The renormalization flow is shown in Fig.11.
The Kosterlitz-Thouless transition is a beautiful transition that
occurs in two dimensions.  It was proposed that the transition was
associated with the unbinding of vortices, that is, the K-T transition
is a transition of the binding-unbinding transition of vortices.

The Kondo problem is also described by the same equation.
In the s-d model, we put
\begin{equation}
x=\pi\beta J_z-2,~~~ y=2|J_{\perp}|\tau,
\end{equation}
where $J_z$ and $J_{\perp}(=J_x=J_y)$ are exchange coupling constants
between the conduction electrons and the localized spin, and $\beta$
is the inverse temperature.  $\tau$ is a small cutoff with
$\tau\propto 1/\mu$.  The scaling equations for the s-d model 
are\cite{kon12,and70,and69,yuv70}
\begin{eqnarray}
\tau\frac{\partial x}{\partial\tau}&=& -\frac{1}{2}y^2,\\
\tau\frac{\partial y}{\partial\tau}&=& -\frac{1}{2}xy.
\end{eqnarray}
The Kondo effect occurs as a crossover from weakly correlated region
to strongly correlated region.  A crossover from weakly to
strongly coupled systems is a universal and ubiquitous phenomenon
in the world.  There appears a universal logarithmic anomaly
as the result of the crossover.
QCD also belogs to the same universality class.
The essence of the K-T transition is that this transition is
a crossover from the weak coupling region to the strong coupling
region.

\vspace{0.5cm}
\begin{figure}[ht]
\includegraphics[width=5.8cm]{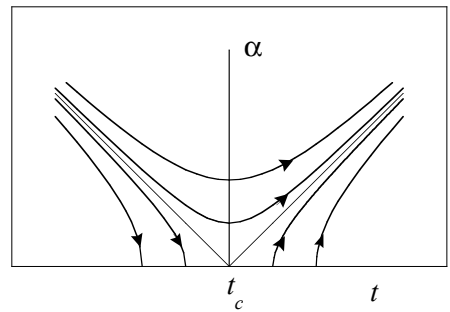}
 \caption{
The renormalization group flow for the sine-Gordon model
as $\mu\rightarrow\infty$..
 }
\end{figure}

\section{Scalar Quantum Electrodynamics}

We have examined the $\phi^4$ theory and showed that there is a
phase transition.  This is a second-order transition.
What will happen when a scalar field couples with electromagnetic
field?  This issue concerns the theory of a complex scalar field $\phi$
interacting with the electromagnetic field $A_{\mu}$, called the
scalar quantum electrodynamics (scalar QED).  The Lagrangian is
\begin{equation}
\mathcal{L}= \frac{1}{2}\big|D_{\mu}\phi\big|^2
-\frac{1}{4}g(|\phi|^2)^2-\frac{1}{4}F_{\mu\nu}^2,
\end{equation}
where $g$ is the coupling constant and 
$F_{\mu\nu}=\partial_{\mu}A_{\nu}-\partial_{\nu}A_{\mu}$.
$D_{\mu}$ is the covariant derivative given as
\begin{equation}
D_{\mu}= \partial_{\mu}-ieA_{\mu},
\end{equation}
with the charge $e$.  The scalar field $\phi$ is an $N$ component
complex scalar field: $\phi=(\phi_1,\cdots,\phi_N)$.
This model is indeed a model of a superconductor.  The renormalization
group analysis shows that this model exhibits a first-order transition
near four dimensions $d=4-\epsilon$ when $2N<365$\cite{col73,hal74,hik79,lub78,che78}.
Coleman and Weinberg first considered the scalar QED model in the
case $N=1$.  They called this transition the dimensional transmutation.
This means that the scalar field theory acquires a mass as the rresult
of radiative corrections.  The result based on the $\epsilon$-expansion
predicts that a superconducting transition in a magnetic field is a
first-order transition.  This transition may be related to a first-order
transition in a magnetic field al low temperatures\cite{mak64}.

The bare and renormalized fields and coupling constants are
defined as
\begin{eqnarray}
\phi_0 &=& \sqrt{Z_{\phi}}\phi, \\
g_0 &=& \frac{Z_4}{Z_{\phi}^2}g\mu^{4-d}, \\
e_0 &=& \frac{Z_e}{\sqrt{Z_AZ_{\phi}}}e, \\
A_{\mu0} &=& \sqrt{Z_A}A_{\mu},
\end{eqnarray}
where $\phi$, $g$, $e$ and $A_{\mu}$ are renormalized quantities.
We have four renormalization constants.  Thanks to the Ward identity
\begin{equation}
Z_e= Z_{\phi},
\end{equation}
three renormalization constants should be determined.
We show the results:
\begin{eqnarray}
Z_{\phi} &=& 1+\frac{3}{8\pi^2\epsilon}, \\
Z_A &=& 1-\frac{2N}{48\pi^2\epsilon}e^2, \\
Z_g &=& 1+\frac{2N+8}{8\pi^2\epsilon}+\frac{3}{8\pi^2\epsilon}\frac{1}{g}e^4 .
\end{eqnarray}
The renormalization group equations are given by
\begin{eqnarray}
\mu\frac{\partial e^2}{\partial\mu} &=& -\epsilon e^2
+\frac{N}{24\pi^2}e^4 , \\
\mu\frac{\partial g}{\partial\mu} &=& -\epsilon g+\frac{N*4}{4\pi^2}g^2
+\frac{3}{8\pi^2}e^4-\frac{3}{4\pi^2}e^2g.
\end{eqnarray}
The fixed point is given by
\begin{eqnarray}
e_c &=& \frac{24}{N}\pi^2\epsilon , \\
g_c &=& \epsilon\frac{2\pi^2}{N+4}\Big[ 1+\frac{18}{N}
\pm\frac{\sqrt{n^2-360n-2160}}{n}\Big] ,
\end{eqnarray}
where $n=2N$.  The square root $\delta\equiv\sqrt{n^2-360n-2160}$ is
real when $2N>365$.  This indicates that the zero of a set of beta 
functions exists when $N$ is sufficiently large as large as $2N>365$.
Hence there is no continuous transition when $N$ is small, $2N<365$,
and the phase transition is first order.

There are also calculations up to the two-loop order for scalar
QED\cite{kol90,fol96}.  This model is also closely related with the 
phase transition from a smectic-A to a nematic liquid crystal for
which a second-order transition was reported\cite{dav79}.
When $N$ is large as far as $2N>365$, the transition becomes second
order.  Does the renormalization group result for the scalar QED
contradict with second-order transition in superconductors?
This subject has not been solved yet.  A possibility of second-order
transition was investigated in three dimensions by using the
renormalization group theory\cite{her96}.  An extra parameter $c$
was introduced in \cite{her96} to impose a relation between
the external momentum $p$ and the momentum $q$ of the gauge
field as $q=p/c$.  It was shown that when $c>5.7$, we have a 
second-order transition.  We don't think that it is clear whether
the introduction of $c$ is justified or not.

\section{Summary}

We presented the renormalization group procedure for several
important models in field theory on the basis of the dimensional
regularization method.  The dimensional method is very useful
and the divergence is separated from an integral without ambiguity.
We investigated three fundamental models in field theory:
$\phi^4$ theory, non-linear sigma model and sine-Gordon model.
These models are often regarded as an effective model in
understanding physical phenomena.  The renormalization group
equations were derived in a standard way by regularizing the
ultraviolet divergence.  The renormalization group theory is
useful in the study of various quantum systems.

The renormalization means that the divergence, appearing in
the evaluation of physical quantities, are removed by 
introducing the finite number of renormalization constants.
If we need infinite number of constants to cancel the divergences
for some model, that model is called nonrenormalizable.
There are many renormalizable field theory models.
We considered three models among them.  The idea of 
renormalization group theory arises naturally from
renormalization.  The dependence of physical quantities on
the renormalization energy scale easily leads us to the of
renormalization group.

\section{acknowledgment}
I express my sincere thanks to Prof S. Hikami
(Okinawa Institute for Science and Technology).
His lecture on the renormalization group theory was very enlightening
and I gained a great deal of benefit from his lecture.

\end{document}